\documentclass[aps,pra,preprint,groupedaddress,showpacs]{revtex4-1}
\usepackage{graphicx}
\usepackage{dcolumn}
\usepackage{bm}
\begin{document}

\title{Compressive adaptive computational ghost imaging}

\author{Marc A\ss mann$^{\ast}$}
\author{Manfred Bayer}
\affiliation{Experimentelle Physik 2, Technische Universit\"at
Dortmund, 44221 Dortmund, Germany}

\date{\today}

\begin{abstract} Compressive sensing is considered a huge breakthrough in signal
acquisition. It allows recording an image consisting of $N^2$ pixels
using much fewer than $N^2$ measurements if it can be transformed to
a basis where most pixels take on negligibly small values. Standard
compressive sensing techniques suffer from the computational
overhead needed to reconstruct an image with typical computation
times between hours and days and are thus not optimal for
applications in physics and spectroscopy. We demonstrate an adaptive
compressive sampling technique that performs measurements directly
in a sparse basis. It needs much fewer than $N^2$ measurements
without any computational overhead, so the result is available
instantly.
\end{abstract}

\maketitle

Computational ghost imaging (CGI) is a novel imaging technique that
has received significant attention during the last few
years\cite{Erkmen2010}. It is a consequent further development of
conventional ghost imaging \cite{Pittman1995,Ferri2010} which allows
to record spatially resolved images using a detector without spatial
resolution. In conventional ghost imaging the image is recorded
using two spatially correlated light beams, one object and one
reference beam. The object beam illuminates the object to be imaged
and is then collected using a bucket detector. The reference beam
never interacts with the object and is recorded using a pixelated
device offering spatial resolution. As both beams are spatially
correlated the coincidence count signal allows one to retrieve a
ghost image of the object. Ghost imaging using both entangled
photons \cite{Pittman1995} or classical light
\cite{Bennink2002,Valencia2005,Cheng2004} as the spatially
correlated twin beam source has been demonstrated. A seminal paper
by Shapiro \cite{Shapiro2008} clarified that the sole purpose of the
reference beam lies in determining the illumination pattern at the
object position at each instant, while the object beam gives data
about the transmission of this pattern through the object.
Therefore, if one can create a deterministic illumination pattern at
the object position, the reference beam becomes obsolete and CGI
using just a single beam and a single pixel detector
\cite{Duarte2008} becomes possible. This approach has been realized
using deterministic speckle patterns created using a spatial light
modulator (SLM) \cite{Bromberg2009}. It has also been demonstrated
that this technique also offers the possibility to perform
compressive sensing \cite{Katz2009,Zerom2011}, that is it is
possible to reconstruct an image consisting of $N^2$ pixels using
much less than $N^2$ measurements by utilizing the fact that natural
images are typically sparse\cite{Candes2008}: When transformed to an
appropriate basis, they contain many coefficients that are zero or
close to it. In practice, the transmission measured for each speckle
pattern constitutes a projection of the object image and compressive
sensing is performed by utilizing an algorithm which checks all the
possible images which are consistent with the projections performed
and finds the image which is the sparsest one. Usually the $L1$-norm
serves as a measure of sparsity and the image which minimizes it, is
the optimal reconstruction of the object. However, this method still
has some drawbacks. The time taken by the reconstruction algorithm
can become very long for large images and one needs to compute the
speckle pattern at the object position by using the Fresnel-Huygens
propagator on the phase pattern imprinted on the SLM. While the
latter is not a big problem - one can calculate the speckle pattern
once and reuse the phase pattern masks - the computational overhead,
given by the computational effort once all measurements have been
made, is a huge problem. The overhead becomes especially problematic
considering typical problems in spectroscopy (e.g. pump-probe
spectroscopy), where many similar images need to be taken, while one
experimental parameter is changed. Here, it is desirable to have the
reconstructed image directly, so one can use this information when
taking the next image. For example one could adaptively scan the
previous image for regions of large values or strongly varying
values and scan these areas with higher resolution in the next
image.

\section*{results}
\subsection*{The adaptive compressive CGI algorithm} 
\begin{figure}
  \centerline{\includegraphics[width=0.95\linewidth]{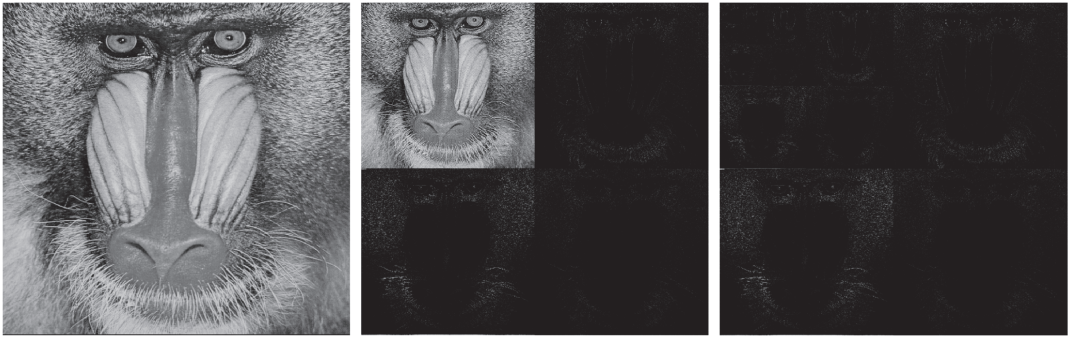}}
  \caption{512$\times$512 pixel baboon test image (left panel) and
  its one-step (middle panel) and complete (right panel) wavelet transform. For the
  transform absolute values of the wavelet coefficients are shown.
  White regions correspond to large wavelet values and mark regions with strong edges.
  Every wavelet coefficient at scale $j$ contains information about four pixels of the
  coarse image of size $\frac{N}{j}\times\frac{N}{j}$. Also, every
  wavelet coefficient has four children wavelet coefficients at
  scale $j-1$ which contain information about the same range of the
  image.}
  \label{Monkey}
\end{figure}

We demonstrate an
alternative way to perform compressive CGI (CCGI) without any
computational overhead once all measurements have been performed by
using an adaptive measurement scheme. We follow a promising strategy
for adaptive compressive sensing that suggests replacing the random
speckle patterns by directly using the patterns that form the sparse
basis \cite{Averbuch2012}. We start the discussion of our strategy
by recalling the properties of the 2D Haar wavelet transform of
square images consisting of $N\times N$ pixels. The wavelet
decomposition procedure is schematically depicted in figure
\ref{Monkey}. The decomposition of the image $I(x,y)$ is performed
seperately for rows and columns. At first each row is divided into
$\frac{N}{2}$ pairs of adjacent pixels. The partial wavelet
transform $T'(x,y)$ now consists of the sum and the difference of
these adjacent pixels according to the following rules for
$x<\frac{N}{2}$:
\begin{eqnarray}
     T'(x,y) & = & I(2x,y)+I(2x+1,y) \\
  T'(x+\frac{N}{2},y) & = & I(2x,y)-I(2x+1,y).
\end{eqnarray}
Repeating that procedure for each column  in $T'$ according to
similar rules for $y<\frac{N}{2}$ gives the full transform $T(x,y)$:
\begin{eqnarray}
     T(x,y) & = & T'(x,2y)+T'(x,2y+1) \\
  T(x,y+\frac{N}{2}) & = & T'(x,2y)-T'(x,2y+1).
\end{eqnarray}
The resulting transform now consists of four quadrants. The upper
left quadrant represents a coarse version of the original image,
while the other three quadrants contain information about
horizontal, vertical and diagonal edges. One may now continue and
perform another wavelet transform on the upper left quadrant and
iteratively repeat this procedure until the pixel in the upper left
corner contains the mean intensity of the picture and all other
pixels contain information about edges. Now each additional
transform performed corresponds to a coarser scale $j$ with wavelet
coefficients spanning over larger regions, but carrying information
over a smaller range of frequencies. Such wavelet representations
are efficient in terms of image compression. Only a small portion of
natural images consists of edges and only wavelet coefficients
corresponding to regions with sharp edges are large, therefore only
few large coefficients are sufficient to approximate the full image.
As can be seen in figure \ref{Monkey}, the number of large wavelet
coefficients (shown in white) is rather small.

\begin{figure}
  \centerline{\includegraphics[width=0.6\linewidth]{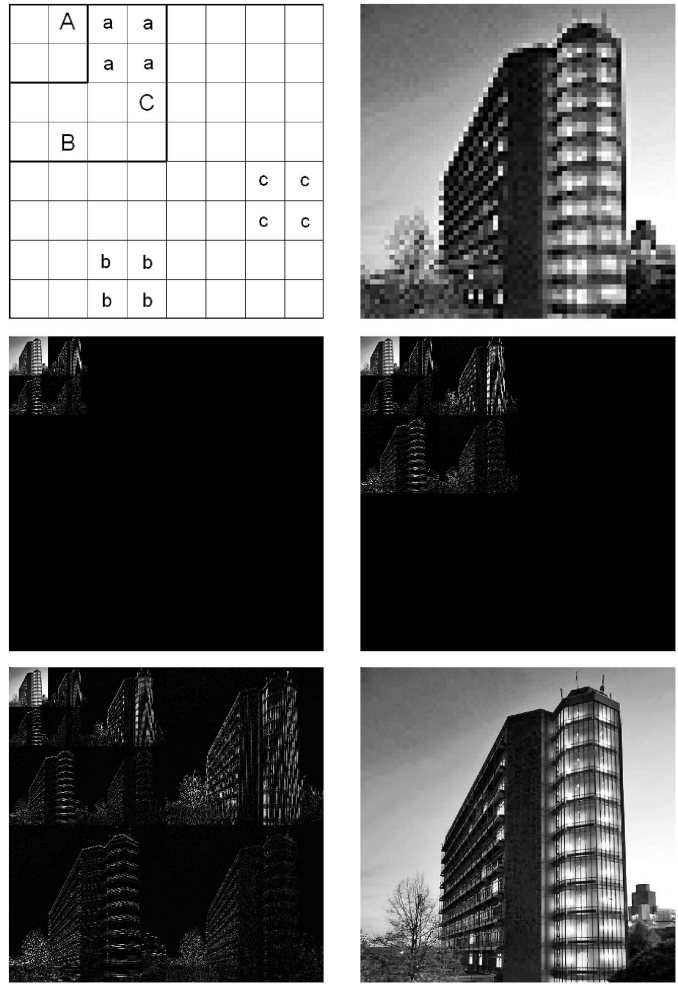}}
  \caption{Summary of the CCGI scheme: First, a low resolution real space image is taken
  (upper right panel). The wavelet transform of that image
  is created (middle left panel). Large wavelet coefficients are shown in white,
  small ones in black. For each wavelet coefficient larger than the chosen threshold,
  its four children coefficients are determined. See the upper left panel for exemplaric
  parent (capital letters) and corresponding children wavelets
  (corresponding lower case letters) across different scales. The measurement of a
  children wavelet coefficient requires four real space measurements at a finer scale.
  After all target wavelet coefficients at the finer scale are measured (middle right panel), the
  procedure continues with the next finer scale until the target scale $j$=1 is reached or
  no wavelet coefficient is larger than the threshold value (lower left panel). The result
  is then converted back to a real space image using the inverse wavelet transform (lower
  right image). For this example the number of measurements needed is roughly 40\,$\%$ of
  the number of pixels present in the image. Note that the upper right, lower left and
  lower right sector of the wavelet transform correspond to horizontal, vertical and
  diagonal edges, respectively. Wavelet coefficients have been multiplied by 8 to
  enhance contrast.}
  \label{example}
\end{figure}

This strategy becomes interesting as the wavelet transformation is
hierarchic. Every parent coefficient at some coarse scale has four
children coefficients at the next finer scale covering the same
spatial region. As it is very likely that the children wavelet
coefficients belonging to parent coefficients which are small will
also be small, this offers a great opportunity for image compression
in terms of wavelet trees \cite{Shapiro1993} by cutting of these
trees at an adequate scale. We follow a similar strategy and first
take a coarse image of size $\frac{N}{j}\times \frac{N}{j}$.
Experimentally, this is realized by inserting a phase-only SLM
(Holoeye-Pluto) in the path of a laser beam polarized such that the
SLM only introduces a phase shift to it. The phase pattern imprinted
on the SLM is the Fourier transform of a square superposed with the
phase map of a lens. As a consequence, in the focal plane behind the
SLM the square is recovered in the spatial intensity pattern of the
light beam. We precomputed 87040 of such phase patterns using an
iterative numerical technique based on the adaptive-additive
algorithm \cite{Dufresne2001}. 65536 of these form the pixels of a
256$\times$256 ($j$=1) pixel square. The other patterns form the
pixels of squares of the same size, but consisting of fewer
(128$\times$128 ($j$=2), 64$\times$64 ($j$=3) and 32$\times$32
($j$=4)), but larger pixels of size $2^{(2(j-1))}$, respectively.
The object to be imaged is placed at the focal plane of the SLM
($f$=36\,cm) and the transmission through that object is measured.
Under the conditions used throughout the manuscript, the whole
square has a side length of 32\,mm. When the coarse image is taken,
we perform a one-step wavelet transform on it. Now we check the
absolute values of the wavelet coefficients corresponding to edges
against a predefined threshold $I_j$. If the values are larger than
$I_j$, the four children wavelet values at the next finer scale
$j$-1 are measured too. As each wavelet coefficient spans over
exactly four pixels at its scale, it is never necessary to perform
more than four measurements in real space to determine any wavelet
value. Once all the measurements at the finer scale have been
performed, a new finer image can be constructed. It consists of the
newly measured transmission values for regions containing sharp
edges and of the transmission values already measured at a coarser
scale for regions without edges. Now another one-step wavelet
transform is performed on this finer image and again all wavelet
values are checked against a new threshold $I_{j-1}$. This process
is repeated until the finest picture at scale $j$=1 is constructed.
A summary of the imaging process is presented in fig. \ref{example}.

\subsection*{Experimental results} 
\begin{figure}
  \centerline{\includegraphics[width=0.95\linewidth]{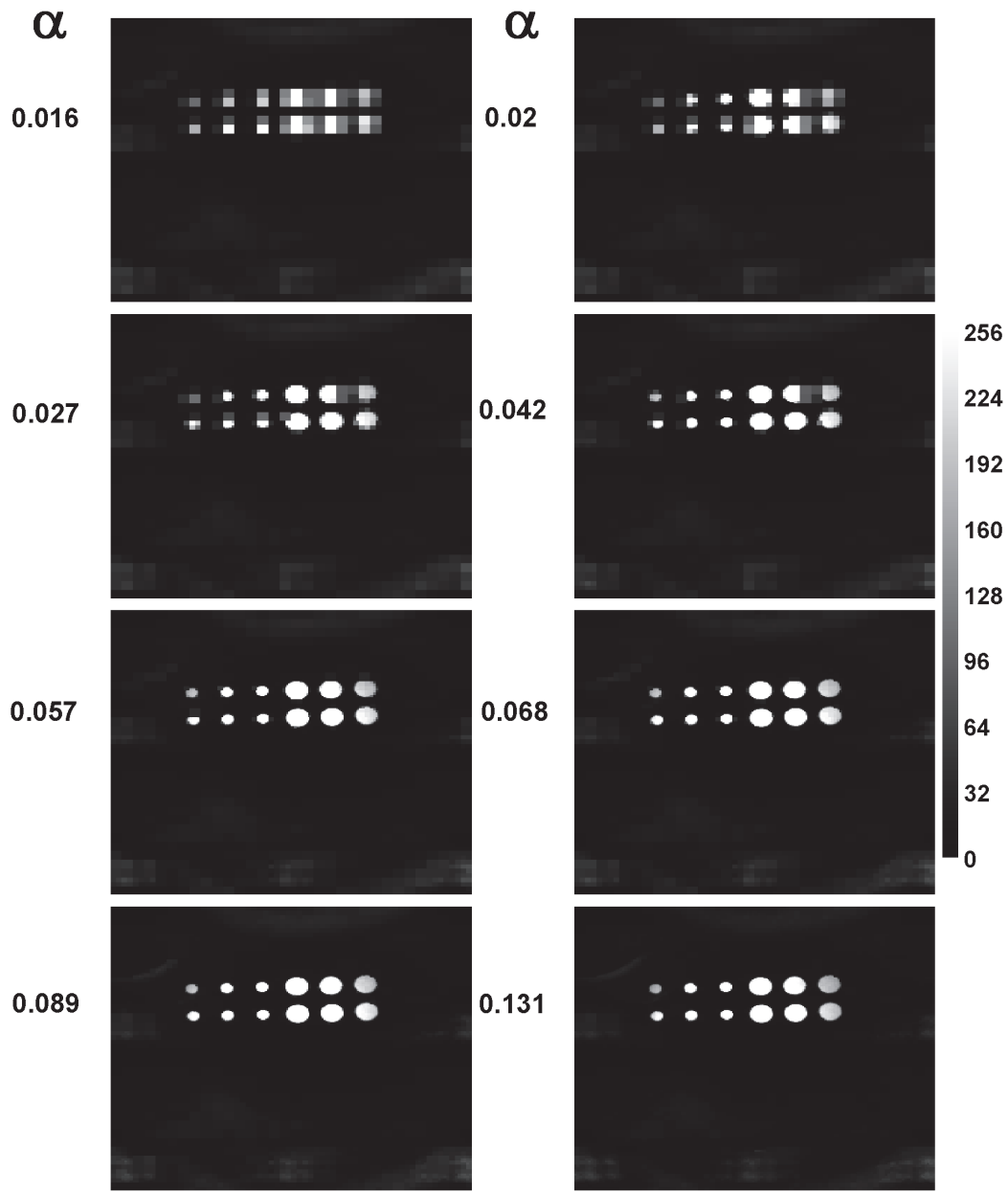}}
  \caption{Normalized transmission maps through a metal plate containing
  twelve holes. The large holes have a diameter of 2\,mm, while the smaller
  ones have a diameter of 1\,mm. $\alpha$ gives the decreased acquisition
  rate. A faithful image of the plate is already possible with
  approximately 5-7\,$\%$ of the measurements required to record every
  single pixel in full resolution.}
  \label{Lochblende}
\end{figure}

We tested the CCGI algorithm using
a metal plate containing twelve holes as a test target. We chose to
use a threshold which becomes sharper at finer scales
($I_{j-1}$=2$I_j$) and varied the initial threshold $I_4$, resulting
in images of differing quality. The results are shown in figure
\ref{Lochblende}. Here the transmission maps quantized to 256
greyscales are shown in terms of the decreased acquisition rate
$\alpha$, which is the total number of measurements performed on all
scales divided by the total number of pixels present on the finest
scale ($N^2$=65536). The transmission is normalized to the empty
space transmission to account for possible inhomogeneities
introduced by the SLM.
As can be seen, the image is reproduced quite well at relatively
small $\alpha$. At around 2\,$\%$ the quality is already sufficient
for distinguishing the holes and counting their number. For $\alpha$
around 4\,$\%$ the image already looks reasonable. For $\alpha$
around 7\,$\%$ the recorded image shows good quality. For larger
$\alpha$ only small improvements are seen. However, to get a more
quantitative measure of the recorded image quality, we calculated
the mean squared error
\begin{equation}
\sigma^2=\frac{1}{N^2}\sum_{i,j} [T(x,y)-R(x,y)]^2
\end{equation}
 of the measured image as compared to the reference construction
 drawing $R(x,y)$ of the metal plate containing the holes.
\begin{figure}
  \centerline{\includegraphics[width=0.95\linewidth]{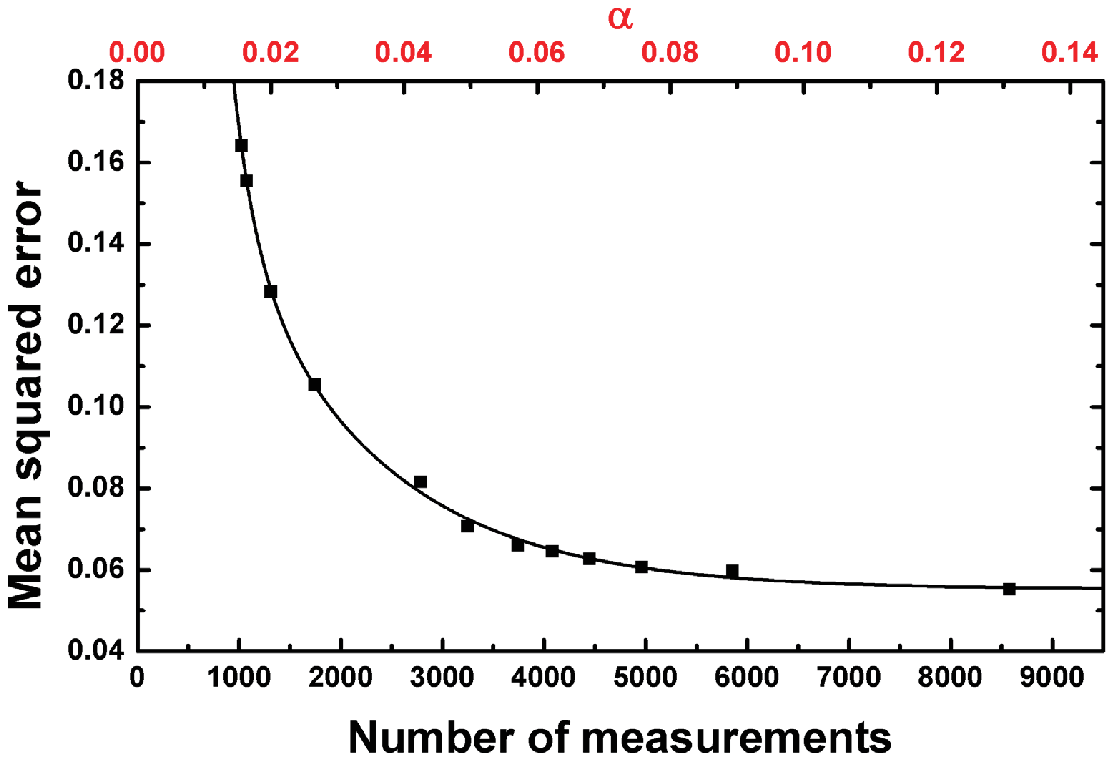}}
  \caption{Mean squared error versus the number of measurements taken.
  The residual error saturates for $\alpha>$7\,$\%$.}
  \label{MSE}
\end{figure}
The impression that the image quality does not improve significantly
for $\alpha >$7\,$\%$ is verified. The mean squared error roughly
follows an exponential decay and saturates approximately at a value
of 0.055 for large $\alpha$. A closer examination of this residual
error shows that it is mainly caused by the edges of the holes. In
contrast to the construction drawing, the edges between full
transmission and zero transmission are not positioned at pixel
borders. Therefore the pixels at the edges show some intermediate
transmission and introduce some deviations from the reference image.
\begin{figure}
  \centerline{\includegraphics[width=0.95\linewidth]{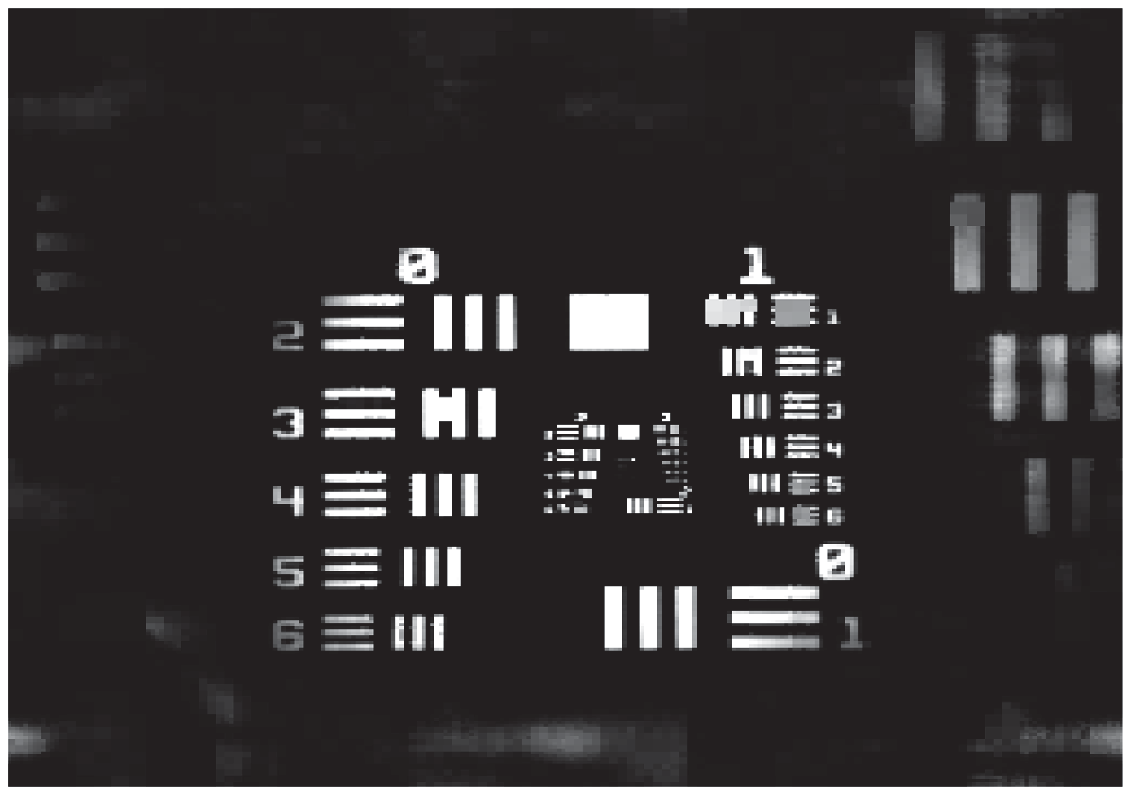}}
  \caption{Normalized transmission map through a 1951 USAF resolution
  test chart at $\alpha$=0.369.}
  \label{USAF}
\end{figure}
The number of necessary measurements needed for near optimal
reconstruction of an image obviously depends on the number of large
wavelet coefficients that image carries. In order to demonstrate the
adaptivity of our technique, we kept the threshold setting used for
measuring the metal plate at $\alpha$=0.131 in figure
\ref{Lochblende} and imaged a more complex object - a 1951 USAF
resolution test chart. The recorded image is shown in figure
\ref{USAF}. The image quality is still good, but the algorithm
automatically took almost three times more measurements than were
needed for the metal plate, resulting in $\alpha$=36.9\,$\%$. The
image resolution is reasonable. One pixel has a length of about
125\,$\mu$m in real space which is roughly the size of the lines on
the test plate which can still be resolved. Nevertheless the image
shows some artifacts. These are a consequence of a weakness of the
algorithm used. Strictly periodic structures like several parallel
lines placed next to each other may look like having no edge at all
at a coarser scale. However, such problems may be overcome by more
advanced algorithms relying on taking more than just the parent
wavelet value into account when deciding on whether a certain
wavelet value should be measured or not \cite{Averbuch2012}.\\Our
technique offers a wide range of advantages. As it is adaptive, one
has control over the quality of the image in advance by choosing the
thresholds. The algorithm does not require any additional
computationally intensive recovery algorithms needed for standard
compressive sensing techniques using pseudorandom illumination
patterns. Our technique works reasonable fast. The SLM can be
operated at up to 60\,Hz. Photodiodes and readout circuitry working
on the same timescale are common today, opening up the possibility
to record images of reasonable resolution and quality within few
minutes. In particular experimental techniques requiring single
pixel detectors like lock-in detection for sensitive pump-probe
measurements may benefit from our results. Spatially resolved
measurements are to the best of our knowledge not carried out with
high resolution using such techniques due to the long measurement
durations that would be needed. Reducing the number of necessary
measurements by a factor of at least fifteen opens up the way to
perform such measurements with high spatial resolution. If the
duration of the measurement is more crucial than the image quality,
our approach also allows one to perform a fixed number of finer
measurements for a preset number of largest wavelet coefficients at
each scale instead of using thresholds. In that way it is possible
to take an image in a fixed amount of time.
\section*{Discussion}
In conclusion we have developed an adaptive CCGI technique that
allows us to record images using a single pixel camera at an
acquisition rate fifteen to twenty times below the Shannon limit by
recording the image directly in a sparse basis. A number of further
research directions arise from our work. Compressive imaging
techniques are not limited to recording image information, but have
also found usage far beyond simple imaging applications in fields
like quantum process tomography \cite{Shabani2011,Liu2012}, optical
encryption \cite{Clemente2010}, fluorescence microscopy and
hyperspectral imaging \cite{Studer2012}. Also our approach is quite
generic. Optimized approaches which also take the magnitude of
neighboring wavelet coefficients into account \cite{Averbuch2012}
instead of just the parent wavelet coefficient may lead to increased
image quality or smaller values of $\alpha$. Also, it is well known
that especially designed measurement matrices can drastically reduce
the number of needed measurements for exact image reconstruction
using seeded belief propagation techniques\cite{Krzakala2012}.
Finally, it is not strictly necessary to use precomputed phase
patterns, but one could compute them on the fly, thereby allowing
one to even choose an adaptive wavelet basis. Yet, the greatest
strength of our approach lies in drastically reducing the needed
measurement time for high-resolution images using single-pixel
detectors without having any need for computational image
reconstruction.\\We would like to conclude this paper by a
comparison between CCGI and standard random Gaussian matrix based
compressive sensing techniques (RMCS) to identify the strengths and
weaknesses of our approach in more detail. Obviously, having the
image available once all the measurements are taken, is an
advantage, but it is also introduces a drawback: CCGI needs to use a
predefined sparse basis, while RMCS will automatically find an
adequate sparse basis during reconstruction. Accordingly, it is
typically possible to achieve a near-ideal reconstruction with fewer
measurements using RMCS. However, it should be noted that the exact
number of measurements needed for near-ideal reconstruction is
usually not known a priori as it depends on the sparsity of the
image. The needed number of measurements is also unknown in CCGI,
but specifying the desired image quality in terms of the threshold
$I_j$ ensures that not too many measurements are made. The exact
number of measurements needed for CCGI is hard to predict as it
depends on how sparse an image is in the wavelength basis. For very
sparse images, the penalty can be as much as 50\%. For less sparse
images, the differences are less drastic. However, for a comparable
number of measurements, CCGI-based methods tend to achieve a better
signal to noise ration than RMCS methods do. See \cite{Averbuch2012}
for a detailed comparison of a slightly modified version of CCGI
with state of the art RMCS techniques. Another important benchmark
is the performance of compressive sensing techniques in the presence
of noise. In CCGI noise can become a severe problem if the noise
magnitude becomes comparable to the threshold chosen. CCGI is
therefore not the method of choice for measuring images containing
strong noise or weak signals. Another issue is scalabilty. Going to
larger images, increases the necessary number of measurements a
computations during data acquisition in CCGI and the complexity of
the minimization problem in RMCS. However, for all the image sizes
we examined, the time needed for performing the measurements was
always so much longer than the time needed for performing the
wavelet transforms and building the sampling queue that no delay was
noticeable. In summary, although other compressive sensing methods
based on Gaussian random matrices approaches typically need fewer
measurements than most (but not all \cite{Monajemi2012}) techniques
using deterministic matrices, having the result immediately renders
adaptive spatially resolved pump-probe spectroscopy and other
delicate spectroscopic techniques with high resolution possible.
Therefore, we suggest that our technique is well suited for
specialized complex problems in physics and spectroscopy which are a
priori known to be reasonably sparse in the wavelet basis, while
RMCS methods are a much better choice for taking single images, for
images where noise is an issue and for images where the sparsest
basis is unknown.

\section*{Methods}
The objects to be imaged were placed at the focal plane of the SLM
($f$=36\,cm) and the transmission through the objects was measured
by a standard commercial photo diode onto which the transmission
through the object was focused and a Keithley 2000 multimeter was
used for measuring the photo diode output. The hole test plate used
consisted of twelve holes. Six of these holes had a diameter of
2\,mm, while the other six holes had a diameter of 1\,mm. Their
average separation was around 1.5\,mm. The laser used for the
transmission measurements was a pulsed Ti:Sapph laser emitting
pulses wih a duration of approximately 2\,ps at a wavelength of
800\,nm.

\section*{Author Contributions}
M.A. designed the experiment and analyzed the data. M.A. and M.B.
wrote the manuscript.

\section*{Competing financial interests}
The authors declare no competing financial interests.
\end{document}